\documentclass[manuscript]{aastex}

\newif\ifblackandwhite
\blackandwhitefalse
\usepackage{amsmath}
\usepackage{natbib}

\usepackage{longtable}

\slugcomment{}
\shorttitle{Periodic $O-C$ Variations in EC20058-5234}
\shortauthors{Dalessio et. al.}

\begin{document}

\renewcommand{\thefootnote}{\fnsymbol{footnote}}
\title{Periodic Variations in the $O-C$ Diagrams of 5 Pulsation Frequencies of the DB White Dwarf EC 20058-5234 \footnote{Based on observations obtained at the Southern Astrophysical Research (SOAR) telescope, which is a joint project of the Minist\'{e}rio da Ci\^{e}ncia, Tecnologia, e Inova\c{c}\~{a}o (MCTI) da Rep\'{u}blica Federativa do Brasil, the U.S. National Optical Astronomy Observatory (NOAO), the University of North Carolina at Chapel Hill (UNC), and Michigan State University (MSU).}}

\author{J. Dalessio\altaffilmark{1,2}, D. J. Sullivan\altaffilmark{3,4}, J. L. Provencal\altaffilmark{1,2}, H. L. Shipman\altaffilmark{1,2}, T. Sullivan\altaffilmark{3,4}, D. Kilkenny\altaffilmark{5}, L. Fraga\altaffilmark{6}, and R. Sefako\altaffilmark{7}}

\altaffiltext{1}{Department of Physics and Astronomy, University of Delaware, Newark, DE 19716 USA}
\altaffiltext{2}{The Delaware Asteroseismology Research Center, DE, USA}
\altaffiltext{3}{School of Chemical \& Physical Sciences, Victoria University of Wellington, P Box 600, Wellington 6012}
\altaffiltext{4}{Visiting astronomer, Mt John University Observatory, operated by the Department of Physics \& Astronomy, University of Canterbury}
\altaffiltext{5}{Department of Physics, University of the Western Cape, Private Bag X17, Bellville 7535, South Africa}
\altaffiltext{6}{Southern Observatory for Astrophysical Research, Casilla 603, La Serena, Chile}
\altaffiltext{7}{South African Astronomical Observatory, PO Box 9, Observatory 7935, South Africa}

\begin{abstract}
Variations in the pulsation arrival time of five independent pulsation frequencies of the DB white dwarf EC 20058-5234 individually imitate the effects of reflex motion induced by a planet or companion but are inconsistent when considered in unison. The pulsation frequencies vary periodically in a $12.9$ year cycle and undergo secular changes that are inconsistent with simple neutrino plus photon cooling models. The magnitude of the periodic and secular variations increase with the period of the pulsations, possibly hinting that the corresponding physical mechanism is located near the surface of the star. The phase of the periodic variations appears coupled to the sign of the secular variations. The standards for pulsation timing based detection of planetary companions around pulsating white dwarfs, and possibly other variables such as subdwarf B stars, should be reevaluated. The physical mechanism responsible for this surprising result may involve a redistribution of angular momentum or a magnetic cycle. Additionally, variations in a supposed combination frequency are shown to match the sum of the variations of the parent frequencies to remarkable precision, an expected but unprecedented confirmation of theoretical predictions.
\end{abstract}

%\keywords{}

\section{Introduction and Formalism}\label{ocsect}

The orbital motion of a planet hosting star about the system's mass center, often referred to as a ``wobble'', will cause variations in the time it takes for light from the star to reach external observers. If the star provides some predictable repetitive behavior, like regular pulsations, the wobble will cause systematic differences between the time the behavior is predicted to be observed and when it is actually observed. This phenomenon has resulted in the discovery of the first known exoplanet and the rest of the pulsar planets \citep{1992Natur.355..145W,1993ApJ...412L..33T}, confirmation and detection of a growing number of planets with Kepler transit timing variations (see \citealt{2010Sci...330...51H} and subsequent Kepler publications), a planet around a sub-dwarf B star \citep{2007Natur.449..189S} as well as strong detection limits on several pulsating white dwarfs \citep{2008ApJ...676..573M,2010AIPC.1273..446H} and one tentative white dwarf planet candidate \citep{2009ApJ...694..327M}. However, changes to the actual frequency of the repetitive behavior will also manifest as differences between the expected and actual time the behavior is observed. To illustrate this, imagine observations of two distant ticking clocks.  Clock one keeps poor time. Everyday the period of the ticks changes in such a way that clock one appears a second slow at noon and a second fast at midnight. Clock two keeps perfect time but is moving in a light second orbit in the plane of the observer with a period of a day. While these clocks are fundamentally very different, the two clocks are indistinguishable in terms of timing alone. It has been asserted via lex parsimoniae (Occum's Razor), specifically for white dwarf pulsators, that any sinusoidal or ``planet like'' variations in the timing of the pulsations are most likely due to orbital motion as there is no known mechanism that would conspire to make the actual pulsation frequency vary in such a manner. This paper presents conclusive evidence to the contrary: frequency variations in a white dwarf pulsator that can mimic the effect of a planetary companion. 

The dominant cooling mechanism for the hot pulsating DB white dwarfs is thought to be the production of plasmon neutrinos. The frequency of the pulsations are determined in part by the temperature, so cooling should cause a slow secular change in frequency. This effect has been well studied with the far cooler, photon cooling dominated DA white dwarf G117-B15A \citep{2005ApJ...634.1311K}. If the rate of frequency change of a hot pulsating DB white dwarf was measured to sufficient precision, the neutrino production rate could be extracted and used to constrain the Standard Model \citep{2004ApJ...602L.109W}. This paper presents evidence that there are other processes that cause large secular changes to multiple pulsation frequencies in a white dwarf. 

The diagnostic tool used to measure variations in pulsation arrival time and pulsation frequency is called the $O-C$, or ``observed minus calculated'' diagram. Depending on the implementation, ``observed minus calculated'' can be a misnomer. In many cases it is a plot of successive measurements of the absolute phase of a periodic variation assuming constant frequency. The ordinate and abscissa of the $O-C$ diagram are both expressed in units of time, but are sometimes normalized by the typical period of the behavior into units of cycles (often labeled ``epochs'' on the abscissa), radians, or degrees.  The $O-C$ diagram can reveal frequency changes that are difficult to detect by direct measurement. In addition to the detection of planets around pulsars and other pulsating stars, it was the primary diagnostic tool that helped earn Hulse and Taylor earn the Nobel Prize in Physics for their indirect detection of gravitational wave emission \citep{1979Natur.277..437T}. The $O-C$ diagram remains a critical tool across time domain 
astronomy, see e.g. \cite{2012MNRAS.419..959O,2011MNRAS.414.3434B,2008ApJ...676..573M,2012ApJ...757L..21H}.

For pulsating white dwarf stars and other short period variables it is convenient to describe $O-C$ as a time varying absolute phase. Given some fixed observed frequency, $f_{obs}$, the instantaneous amplitude of a pulsation can be rewritten in terms of the time varying absolute phase, $\tau(t)$
\begin{equation}\label{e2}
 H(t) = A \sin (2 \pi f_{obs}(t - \tau(t)) )
\end{equation}

$\tau(t)$ can be mathematically described as (for a complete derivation see \citealt{Dalessio}) 

\begin{equation}
 \boxed{\tau(t) = t_0 - \frac{\delta f_{obs}}{f_{obs}} t - \frac{1}{f_{obs}} \int^t \gamma(t^{\prime}) d t^{\prime} + \frac{q(t)}{c} } \label{main} 
\end{equation}

where $t_0$ is an arbitrary offset depending on choice of $t=0$, $f_{obs}$ is the observed frequency, $\delta f_{obs}$ is the difference between the observed and actual, time averaged, Doppler shifted frequency, $\gamma(t)$ is the perturbation to the frequency, and $q(t)$ is the perturbation to the distance between the object and observer. Note that $\gamma(0)=q(0)=0$.

Except in the case of continuous observation where the number of elapsed cycles is explicitly known, all points in the $O-C$ diagram suffer from the ambiguity that they can be shifted up or down by any integer times the pulsation period. A common assumption is that $O-C$ has been sampled often enough that $\gamma(t)$ and $q(t)$ change slowly and smoothly during the observational gaps. Under this assumption, the location of the points on an $O-C$ diagram can be constrained if there is a single unambiguous configuration that creates a smooth, continuous trend.

\section{Observations}
 EC 20058-5234 \citep{1995MNRAS.277..913K} has been identified as a prime candidate for measurement of the neutrino production rate in a white dwarf (see Section \ref{ocsect}). This motivated many of the over $250,000$ PMT (photo-multiplier tube) and CCD (charge-coupled device) photometric measurements spanning 1994-2011.  EC 20058-5234 was a primary target of the Whole Earth Telescope campaign XCOV15 in 1997 \citep{2008MNRAS.387..137S}, followed by seasonal high speed photometry from the 1m McLellan telescope at Mt. John Observatory and the 36 inch SMARTS telescope at CTIO (see e.g. \citealt{2005ASPC..334..495S}). It was also a tertiary target during XCOV27 and was observed at Magellan in 2003 \citep{2007ASPC..372..629S}. A summary of the observations is shown in Table \ref{obs_table}.

\begin{center}
\begin{longtable}{|cccc|}
\caption{Journal of observations and binning scheme. Note that only the lightcurves from XCOV15 and XCOV27 consist of data from multiple sites.} \label{obs_table} \\

\hline 
   \multicolumn{1}{|c}{\textbf{Observatory}} &
   \multicolumn{1}{c}{\textbf{Date Range}} &
   \multicolumn{1}{c}{\textbf{Instrument}} &
   \multicolumn{1}{c|}{\textbf{Observations}} \\[0.5ex] \hline
   
\endfirsthead

%This is the header for the remaining page(s) of the table...
\multicolumn{4}{c}{{\tablename} \thetable{} -- Continued} \\[0.5ex]
  \hline 
  \multicolumn{1}{|c}{\textbf{Observatory}} &
  \multicolumn{1}{c}{\textbf{Date Range}} &
  \multicolumn{1}{c}{\textbf{Instrument}} &
  \multicolumn{1}{c|}{\textbf{Observations}} \\[0.5ex] \hline
 
\endhead

%This is the footer for all pages except the last page of the table...
\hline
  \multicolumn{3}{l}{{Continued on Next Page\ldots}} \\

\endfoot

%This is the footer for the last page of the table...
  \hline 
\endlastfoot

SAAO & 1994/05/14 - 1994/06/02 & PMT & 5535\\
SAAO & 1994/07/06 - 1994/07/12 & PMT & 6881\\
SAAO & 1994/10/03 - 1994/10/03 & PMT & 1304\\
SAAO & 1994/06/12 - 1994/06/12 & PMT & 1378\\
XCOV15& 1997/07/02 - 1997/07/11 & PMT & 46289\\
Mt. John & 1997/10/05 - 1997/10/06 & PMT & 3162 \\
Mt. John & 1998/07/24 - 1998/07/25 & PMT & 7057\\
Mt. John & 1998/08/14 - 1998/08/14 & CCD & 1153\\
Mt. John & 1999/09/09 - 1999/09/13 & PMT & 1517\\
Mt. John & 2000/07/06 - 2000/07/08 & PMT & 8025\\
Mt. John & 2000/09/05 - 2000/09/05 & PMT & 770\\
Mt. John & 2001/03/30 - 2001/04/01 & PMT & 2101\\
Mt. John & 2001/09/21 - 2001/09/24 & PMT & 7710\\
Mt. John & 2002/04/12 - 2002/04/15 & PMT & 8872\\
Mt. John & 2002/08/01 - 2002/08/06 & PMT & 12672\\
Mt. John & 2002/09/05 - 2002/09/09 & PMT & 3814\\
Magellan & 2003/07/11 - 2003/07/13 & CCD & 2317\\
Magellan & 2003/07/25 - 2003/07/31& CCD & 8865\\
Mt. John & 2003/08/27 - 2003/09/02 & PMT & 12216\\
Mt. John & 2003/09/22 - 2003/09/23 & PMT & 4694\\
Mt. John & 2003/10/31 - 2003/10/31 & PMT & 1229\\
Mt. John & 2004/04/23 - 2004/04/25 & PMT & 2474\\
Mt. John & 2004/05/16 - 2004/05/16 & PMT & 1599\\
Mt. John & 2004/06/09 - 2004/06/16 & PMT & 14554\\
Mt. John & 2004/07/09 - 2004/07/14 & PMT & 17240\\
Mt. John & 2004/08/07 - 2004/08/10 & PMT & 5072\\
CTIO SMARTS & 2004/08/24 - 2004/08/28 & CCD & 2320\\
CTIO SMARTS & 2005/09/21 - 2005/09/25& CCD & 979\\
CTIO SMARTS & 2006/08/31 - 2006/09/13& CCD & 5019\\
CTIO SMARTS & 2007/08/17 - 2007/09/06& CCD & 14117\\
Mt. John & 2007/07/16 - 2007/07/17& PMT & 6281\\
CTIO SMARTS & 2008/08/29 - 2008/09/01 & CCD & 1736\\
Mt. John & 2008/08/30 - 2008/08/31 & CCD & 3092\\
XCOV27\footnote{XCOV27 included observations from SOAR, SAAO, and the CTIO SMARTS .9m.}& 2009/05/18 - 2009/05/26 & CCD & 4609\\
Mt. John & 2010/07/05 - 2010/07/11 & CCD & 20201\\
CTIO SMARTS & 2011/09/08 - 2011/09/15 & CCD & 3584\\

\end{longtable}
\end{center}

\section{Results and Analysis}

\subsection{Data Reduction and Preparation}

The PMT photometry obtained at SAAO in 1994 was reduced as described in \cite{1995MNRAS.277..913K}, while the 1997 XCOV15 photometry and the Mt. John PMT photometry was reduced using the methods outlined in \cite{2008MNRAS.387..137S}. All the Mt. John PMT photometry collected between July 1998 and July 2007 employed a three channel photometer \citep{2000BaltA...9..425S} attached to the one meter McLellan telescope. All CCD image calibration and aperture photometry was performed with Maestro \citep{2010AAS...21545209D}. WQED \citep{2009JPhCS.172a2081T} was used to remove spurious points and to perform the barycentric correction (for a review see \citealt{2010PASP..122..935E}) for the CCD data. The barycentric correction applied to the PMT data were performed using the same implementation as used in WQED \citep{1980A&AS...41....1S}. The photometric measurements were divided into 36 individual lightcurves, one for each observing campaign in Table \ref{obs_table}. The following analysis was also performed by dividing the points into lightcurves for each individual night and observing season. The differences were not significant and do not merit further discussion.

\subsection{Testing for Stability}
The frequency and amplitude of all pulsation frequencies with amplitudes above $0.1\%$ were extracted from two of the largest data sets, XCOV15 in 1997 and the 2007 CTIO SMARTS observations. The results are summarized in Table \ref{stable_table}. The same spectrum of frequencies appears in both years (see the Fourier spectrum from \citealt{2008MNRAS.387..137S}), with the exception that pulsation frequencies F, J, and L have amplitudes below the noise threshold in 2007 and that pulsation frequency E appears to have changed at a statistically significant level. The systematically higher pulsation amplitudes in 2007 are an observational effect due to two companions within several arc seconds of EC 20058-5234 (see the Figure in \citealt{1995MNRAS.277..913K}). The PMT observations included all three objects as it was impractical in the typically $2\arcsec$ seeing conditions at Mt. John to separate flux from the target only. However the CCD data, especially when combined with the better seeing conditions at CTIO and the Magellan site, permitted extraction of just the flux from EC 20058-5234 using synthetic aperture techniques. This effect makes it non-trivial to search for amplitude modulation, which, particularly for RR Lyrae stars, is known to be coupled to phase modulation (the Blazhko effect, see e.g. \citealt{2011ApJ...731...24B}). However, if the pulsation amplitudes are intrinsically constant, the ratio of amplitudes from one set to another should be constant. The ratio of the pulsation amplitudes from 2007 to the pulsation amplitudes in 1997 is shown in the last column of Table \ref{stable_table}. The ratio is consistent with there being no amplitude variation with a reduced $\chi^2$ value of $2.3$ with a statistical likelihood around $10\%$. This is the worst reduced $\chi^2$ value for the model of constant amplitude between any two of the data sets in Table \ref{obs_table}. This value of reduced $\chi^2$ is slightly higher than expected for a constant amplitude. The removal of atmospheric extinction and other observational effects could introduce small changes in observed amplitude so we do not find this slight disagreement particularly alarming. Even if the amplitude of these pulsations frequencies are intrinsically changing, the changes are relatively small and only barely distinguishable between our two best data sets.

\begin{table}\small
\begin{center}
\caption{Pulsation frequencies with amplitudes above $0.1\%$ as directly measured from XCOV15 and the 2007 CTIO SMARTS data sets. The standard error in the measurement of the amplitude of all frequencies was $0.09$ mma (1 mma = 0.1$\%$) in the 2007 CTIO SMARTS dataset and $0.2$ mma in the XCOV15 dataset. Several frequencies from XCOV15 were not detected in the 2007 data set but would have been had the amplitude remained constant. The systematic increase in amplitude from 1997 to 2007 is an observational artifact. The transition from PMTs to CCD imaging which allows removal of some of the flux of the nearby companions. The ratio of pulsation amplitudes from 2007 to 1997 is consistent with having a constant amplitude with a reduced $\chi^2$ value of $2.3$ and a likelihood of around $10\%$.  }\label{stable_table}
\begin{tabular}{|c|c|ll|ll|l|}
\tableline
Label & Period (s) & \multicolumn{2}{c|}{Frequency (uhz)} & \multicolumn{3}{c|}{Amplitude (mma)}\\
 &  (approx) & \multicolumn{1}{c}{1997} &  \multicolumn{1}{c|}{2007} &  \multicolumn{1}{c}{1997} &  \multicolumn{1}{c}{2007} & \multicolumn{1}{c|}{Ratio}\\

\tableline
A & 134 & 7452.18(.05) & 7452.27(.02) & 1.67 & 2.2 & 1.3(.1)\\
B & 195 & 5128.53(.03) & 5128.71(.02) & 2.51 & 3.3 & 1.31(.09)\\
C & 204 & 4902.14(.05) & 4902.14(.02) & 1.47 & 2.6 & 1.8(.2)\\
D & 205 & 4887.84(.03) & 4887.91(.02) & 2.62 & 3.2 & 1.22(.09)\\
E & 257 & 3893.142(.009) & 3893.239(.005) & 8.37 & 11.3 & 1.35(.03)\\
F & 275 & 3640.37(.07) &  & 1.13 & &  \\
G & 281 & 3559.032(.009) & 3559.033(.005) & 8.45 & 10.5 & 1.24(.03)\\
H & 287 & 3489.07(.05) & 3489.05(.02) & 1.58 & 2.2 & 1.4(.1)\\
I & 333 & 2998.67(.03) & 2998.71(.01) & 3.01 & 4.3 & 1.43(.08)\\i
J & 350 & 2852.36(.09) &  & 1.40 &  &  \\
K & 525 & 1903.61(.04) & 1903.49(.02) & 1.83 & 2.44 & 1.3(.1)\\
L & 540 & 1852.53(.04) &  & 1.86 &  &  \\
\tableline
\end{tabular}
\end{center}
\end{table}

\subsection{Extracting Precise Frequencies}
Precise measurements of the pulsation frequencies were made using the ``bootstrapping'' procedure. The process is described as follows. The frequencies above the noise threshold in the July 2004 dataset were extracted. A sum of sinusoids (one sinusoid at each frequency) were fit to the July 2004 dataset using nonlinear least squares and allowing each frequency to converge to the optimum value. After the fit converged, additional data were added to the set, ensuring that each of the frequencies had been measured to sufficient precision so that if the frequency had remained constant, there would be no ambiguity in the number of cycles between any two data points in the combined set. This ensures that the linear term of equation \ref{main} will not dominate the $O-C$ diagram and that the frequencies are clearly separated from the aliases of the combined set. After the new data were added, the frequencies were re-fitted. This process was repeated until most of the 2003 and 2004 data were combined into a single set. The amplitudes of pulsation frequencies F and L were well below the noise threshold and were not considered.  The combined set of all data between September 2003 and August 2004 gave precise enough frequency measurements to merge the entire data set.  However, merging the September 2003 - August 2004 data set with data from August 2003 or 2005 decreased the precision of our frequency measurements. This is an indication that the phase/frequency of the pulsation frequencies are changing by a substantial amount over that time span. Further combination was abandoned beyond the September 2003 - August 2004 pulsation frequency measurements. The results are summarized in Table \ref{bootstrap_table}.

\begin{table}\small
\begin{center}
\caption{Precision of frequency measurements from bootstrapping of data from September 2003 to August 2004. Standard errors are quoted in parenthesis.}\label{bootstrap_table}
\begin{tabular}{|c|c|cc|}
\tableline
Label & Period [s] & Frequency [uhz] & $\frac{1}{\sigma_f}$ [yrs]\\
\tableline
A & 134 & 7452.249(.001) & 32\\
B & 195 & 5128.598(.001) & 32\\
C & 204 & 4902.176(.0009) & 35\\
D & 205 & 4887.8491(.0009) & 35\\
E & 257 & 3893.2482(.0002) & 160\\
G & 281 & 3559.0017(.0003) & 106\\
H & 287 & 3489.056(.001) & 29\\
I & 333 & 2998.7144(.0007) & 45\\
J & 350 & 2852.439(.002) & 19\\
K & 525 & 1903.464(.001) & 24\\

\tableline
\end{tabular}
\end{center}
\end{table}

\subsection{Building the $O-C$ Diagrams}
A sum of sinusoids at the 10 frequencies in Table \ref{bootstrap_table} were linearly fit to each data set. The calculated absolute phase was plotted as $O-C$ for each pulsation frequency. The $O-C$ diagram of all ten pulsation frequencies showed structure that was not consistent with scatter.  It was also apparent that for some frequencies, $O-C$ was changing by a large fraction of the period, and that \textit{none} of the $O-C$ diagrams were consistent with the parabolic trend of a simple neutrino plus photon cooling model \citep{2008CoAst.154...16B}.

It was assumed (see Section \ref{ocsect}) that for each pulsation frequency, $\gamma(t)$ and $q(t)$ change slowly and smoothly during the observational gaps. An unambiguous smooth trend in the $O-C$ diagrams of pulsation frequencies A, D, E, G, H, and I was clearly discernible and the $O-C$ measurements were fixed to an appropriate location on the $O-C$ diagram. Due to the large observational gap between 1994 and 1997, the measurements from 1994 remained unconstrained. The appropriate location in the $O-C$ diagram of the 1994 $O-C$ measurements is further addressed in section \ref{model_section}. No unambiguous trend was apparent in the $O-C$ diagrams of pulsation frequencies C, B, J, and K and they were removed from analysis. The $O-C$ diagrams of pulsation frequencies A, D, E, G, H, and I are shown in Figures \ref{ocA}-\ref{ocI}.

\ifblackandwhite
\else

\begin{figure} 
\includegraphics[width=1\columnwidth,angle=0]{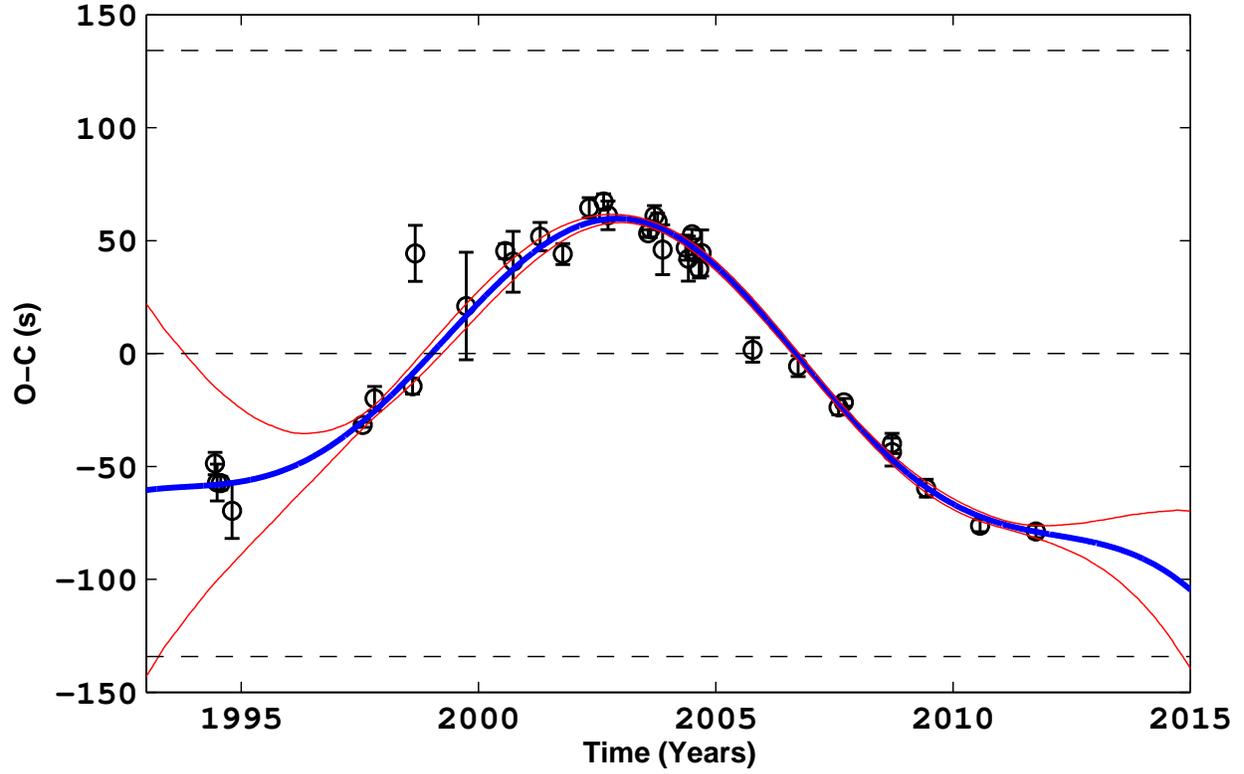}
\caption{$O-C$ of pulsation frequency A. The dashed lines are located at $0$, $P$, and $-P$. The blue line is the best model fit for $\Pi=12.9 yrs$. The red lines indicate the boundaries of the one sigma likelihood prediction of the model not including the 1994 data. Note that a first order polynomial has been removed from the data.}\label{ocA}
\end{figure}
\begin{figure} 
\includegraphics[width=1\columnwidth,angle=0]{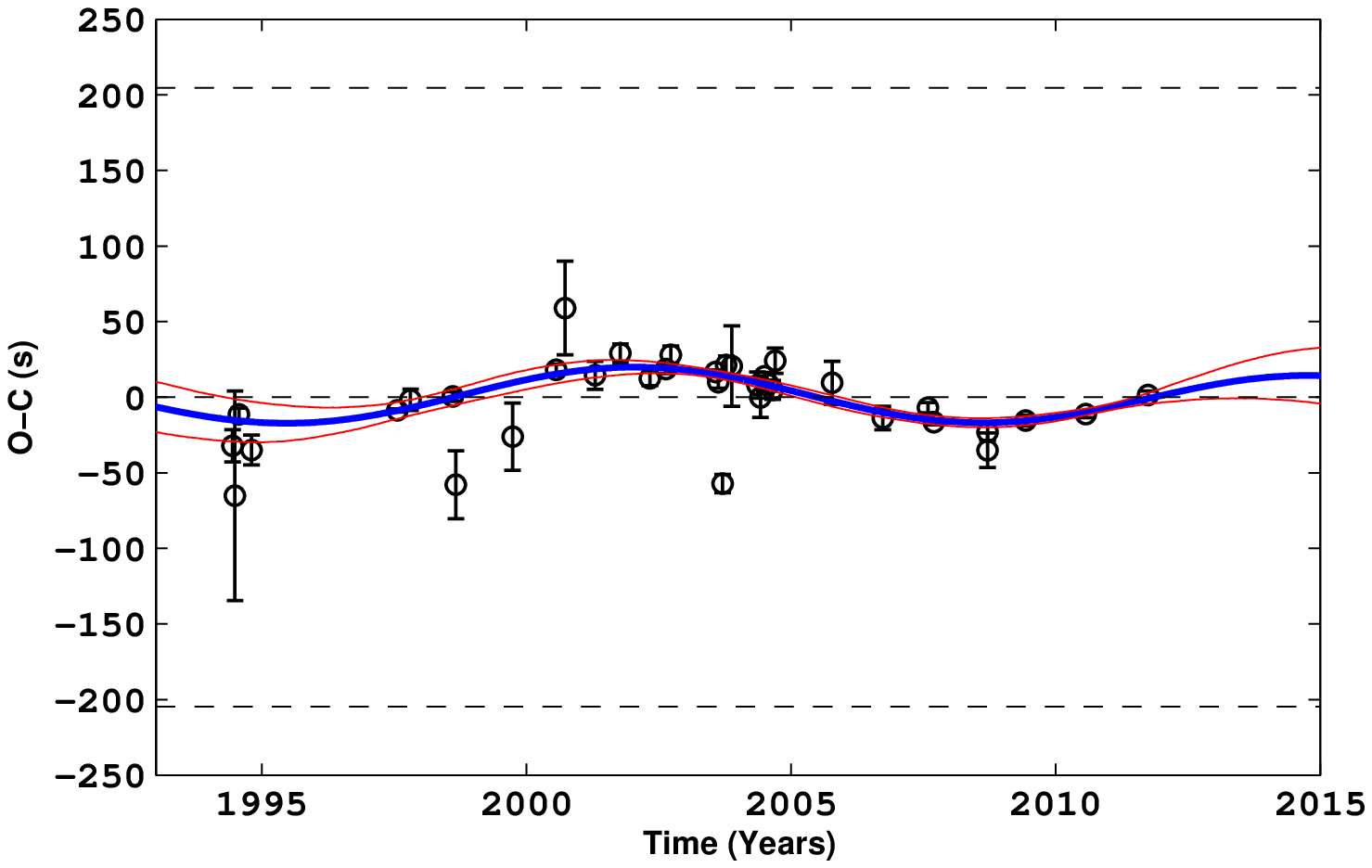}
\caption{$O-C$ of pulsation frequency D. The dashed lines are located at $0$, $P$, and $-P$. The blue line is the best model fit for $\Pi=12.9 yrs$. The red lines indicate the boundaries of the one sigma likelihood prediction of the model not including the 1994 data. Note that a first order polynomial has been removed from the data.}\label{ocD}
\end{figure}
\begin{figure} 
\includegraphics[width=1\columnwidth,angle=0]{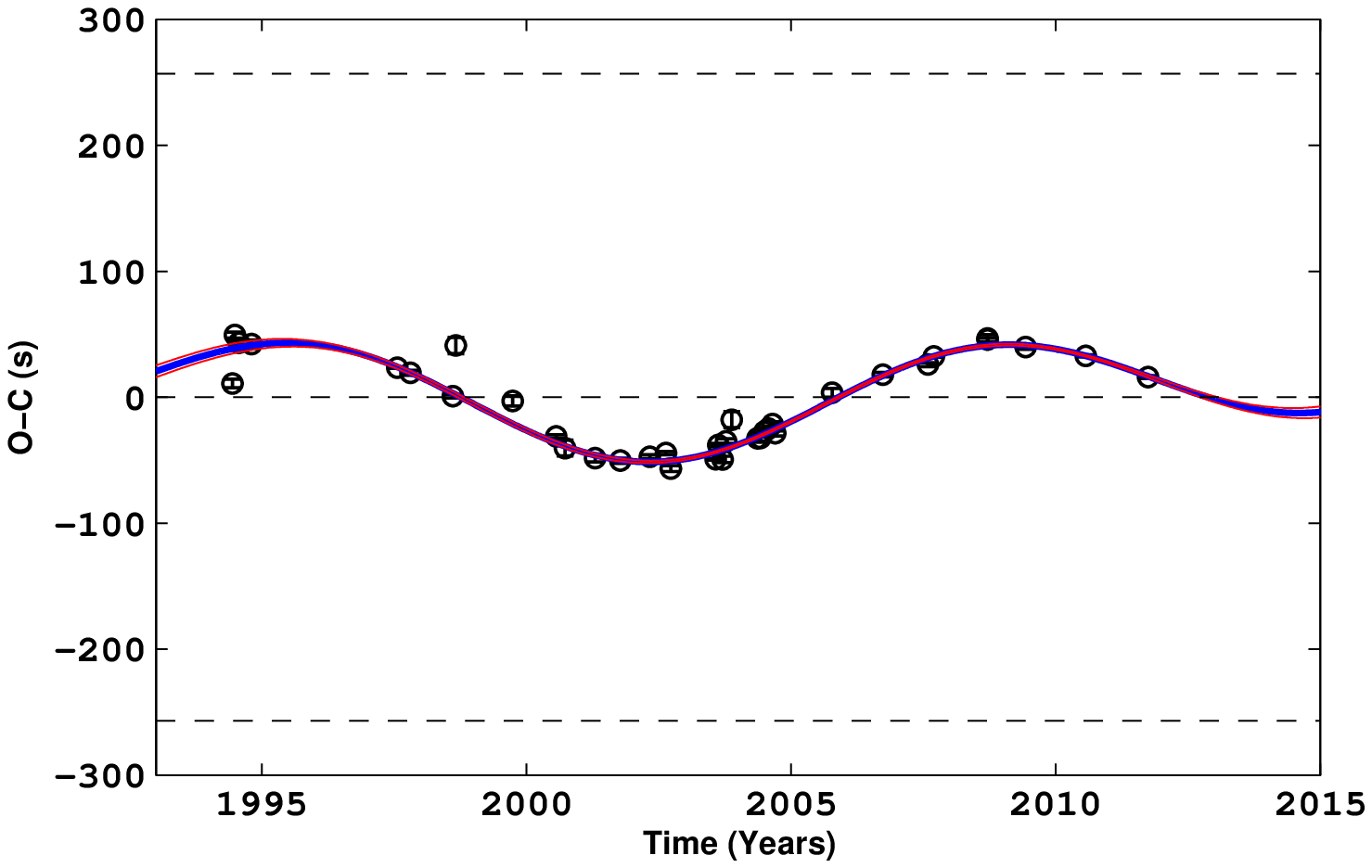}
\caption{$O-C$ of pulsation frequency E. The dashed lines are located at $0$, $P$, and $-P$. The blue line is the best model fit for $\Pi=12.9 yrs$. The red lines indicate the boundaries of the one sigma likelihood prediction of the model not including the 1994 data. Note that a first order polynomial has been removed from the data.}\label{ocE}
\end{figure}
\begin{figure} 
\includegraphics[width=1\columnwidth,angle=0]{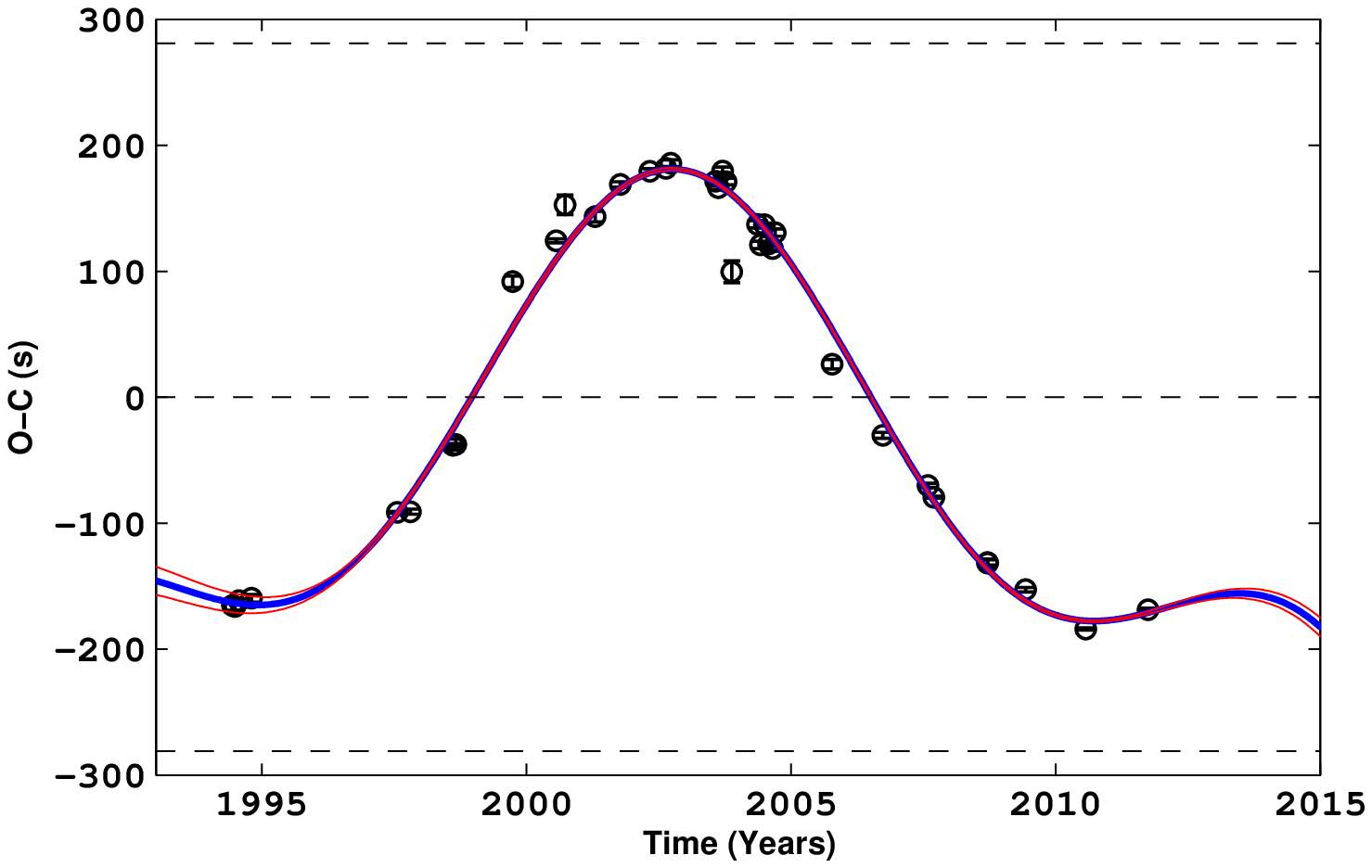}
\caption{$O-C$ of pulsation frequency G. The dashed lines are located at $0$, $P$, and $-P$. The blue line is the best model fit for $\Pi=12.9 yrs$. The red lines indicate the boundaries of the one sigma likelihood prediction of the model not including the 1994 data. Note that a first order polynomial has been removed from the data.}\label{ocG}
\end{figure}
\begin{figure} 
\includegraphics[width=1\columnwidth,angle=0]{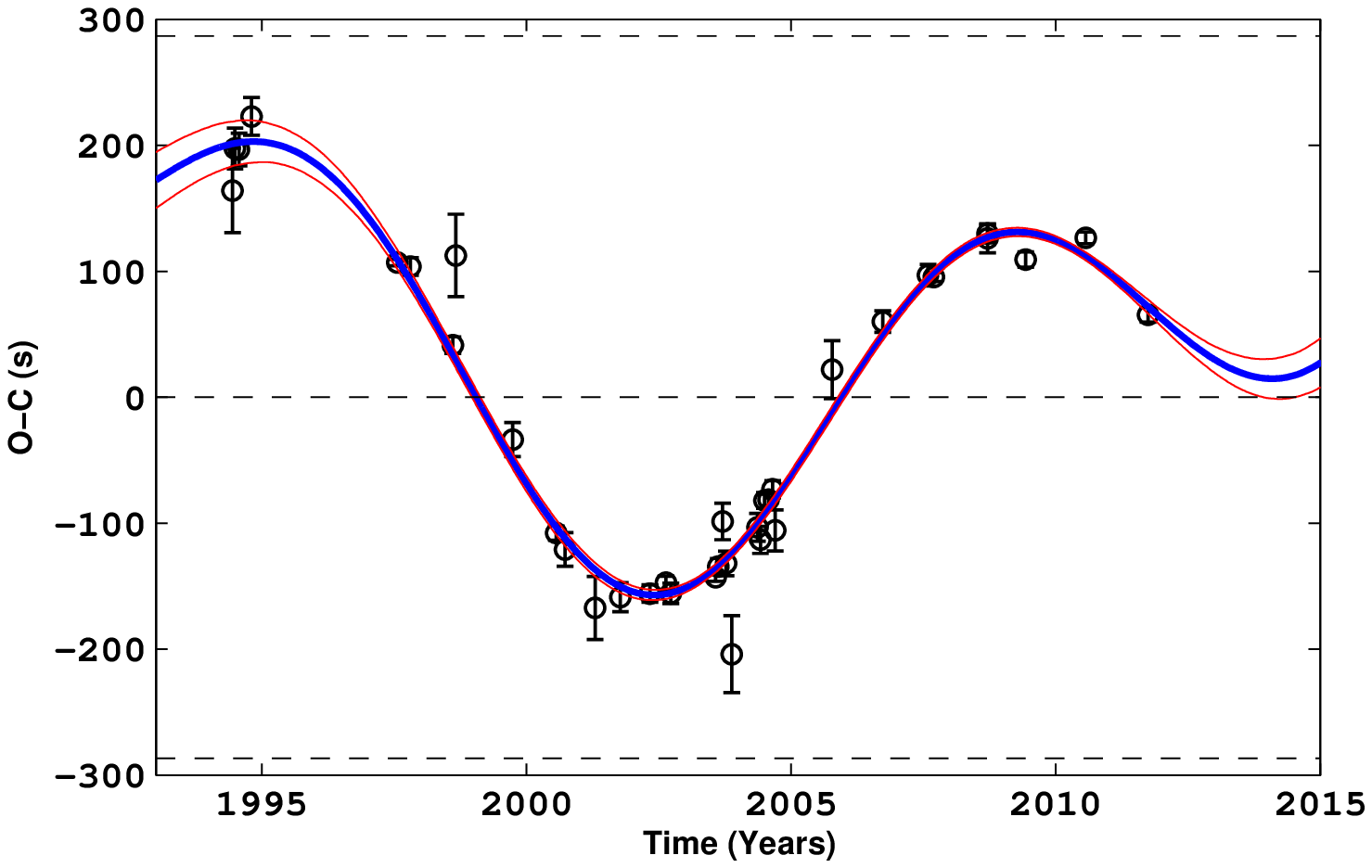}
\caption{$O-C$ of pulsation frequency H. The dashed lines are located at $0$, $P$, and $-P$. The blue line is the best model fit for $\Pi=12.9 yrs$. The red lines indicate the boundaries of the one sigma likelihood prediction of the model not including the 1994 data. Note that a first order polynomial has been removed from the data.}\label{ocH}
\end{figure}
\begin{figure} 
\includegraphics[width=1\columnwidth,angle=0]{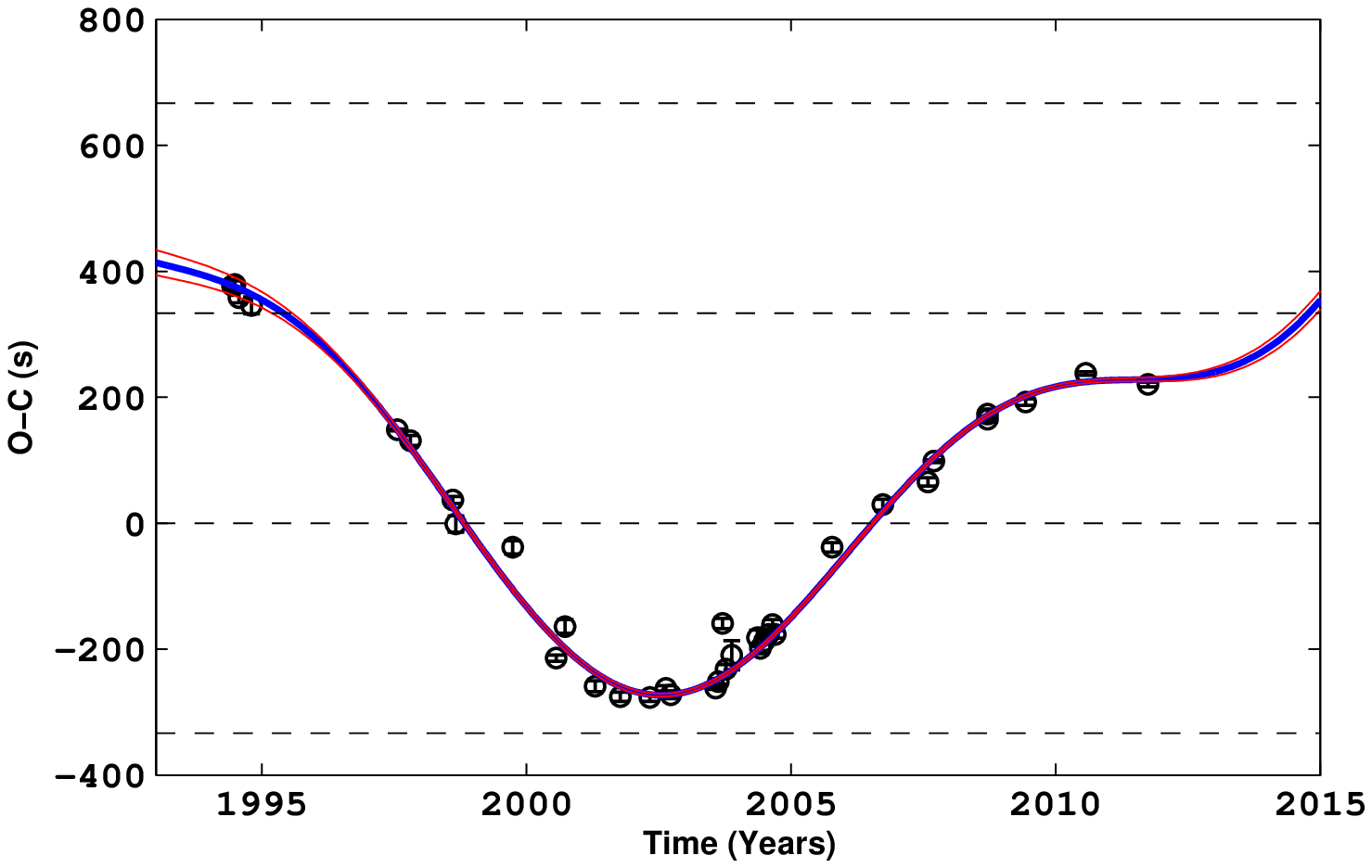}
\caption{$O-C$ of pulsation frequency I. The dashed lines are located at $0$, $P$, $-P$, and $2P$. The blue line is the best model fit for $\Pi=12.9 yrs$. The red lines indicate the boundaries of the one sigma likelihood prediction of the model not including the 1994 data. Note that a first order polynomial has been removed from the data.}\label{ocI}
\end{figure}

\fi

\subsection{Modeling the $O-C$ Variations}\label{model_section}
All six $O-C$ diagrams (Figures \ref{ocA}-\ref{ocI}) show some sort of oscillatory behavior. A sinusoid plus an offset, linear, and parabolic term were chosen as a suitable model. The model is mathematically described as

\begin{equation} \label{model}
\tau(t) = t_0 - \frac{\delta f_{obs}}{f_{obs}} t + \frac{\dot{P}}{2 P_{obs}} t^2  +\frac{\alpha \Pi}{2\pi f_{obs}} \sin \big(\frac{2 \pi}{\Pi}t-\phi \big)
\end{equation}

Instead of fitting this nonlinear model with some initial guesses to $t_0$, $\delta f_{obs}$, the rate of period change $\dot{P}$, the amplitude of the sinusoidal frequency variation $\alpha$, the period of the sinusoidal frequency variation $\Pi$, and the absolute phase of the sinusoidal frequency variation $\phi$, a grid of linear fits can be calculated for a range of $\Pi$. By removing any dependence on initial guesses for the parameters, it is guaranteed that the absolute minimum in $\chi^2$ over the chosen range of $\Pi$ will be found. A Monte Carlo simulation was performed to calculate the statistical likelihood of the possible locations of the 1994 measurements given the model and the 1997-2012 measurements. The one sigma likelihood boundaries are indicated in Figures \ref{ocA}-\ref{ocI}. It was concluded that for all frequencies there was only one reasonable choice for the possible values of the 1994 $O-C$ measurements and that all ambiguity could be considered resolved for the given model. The 1994 $O-C$ measurements were then fixed to these values in each $O-C$ diagram. Fits to Equation \ref{model} were then calculated over a range of $\Pi$ for each of the $O-C$ diagrams. A plot of reduced $\chi^2$ for the fit of each of the six $O-C$ diagrams as a function of $\Pi$ is shown in Figure \ref{pfit}. Four of the six $O-C$ diagrams have a strong periodicity at around 13 years while the other two are not particularly sensitive to the choice of period, but are consistent with a periodicity of 13 years. The minima of the total reduced $\chi^2$ for all six $O-C$ diagrams is at $\Pi = 12.9$ years. This value of $\Pi$ was used to fit each $O-C$ diagram. The best fits are overlayed in Figures \ref{ocA} - \ref{ocI} and Table \ref{fit_table} summarizes the resultant parameters. The large value of reduced $\chi^2$ for some of the fits is disturbing, but the ability of the model to reproduce the \textit{overall} trend in the data is excellent. To further investigate, $\Pi$ was then allowed to vary for each individual $O-C$. This lead to a worse overall reduced $\chi^2$, an indicator that the oscillatory behavior in each $O-C$ is represented appropriately by the same $\Pi$. Fitting the data with high order polynomials did not result in a dramatic improvement in reduced $\chi^2$. An 8th order polynomial, for example, actually increased reduced $\chi^2$ for two of the $O-C$ diagrams and only led to an overall improvement of $7\%$ over Equation \ref{model}. Equation \ref{model} is dramatically more constrained than a polynomial with a similar number of degrees of freedom so this strongly supports the ability of Equation \ref{model} to represent the overall trends in the data. The high values of reduced $\chi^2$ mean that either the error bars in the data are somehow underestimated or there are high frequency variations in $O-C$ that are undersampled. To investigate, the Fourier transforms of the averaged residuals and absolute averaged residuals (see Figure \ref{res}) were calculated. There was no statistically significant or notable power at any frequency, although the residuals appear correlated. It is concluded that there are likely unresolved variations in phase, but these variations do not dilute the effectiveness of Equation \ref{model} in modeling the overall variations.

\ifblackandwhite

\begin{figure} 
\includegraphics[width=1\columnwidth,angle=0]{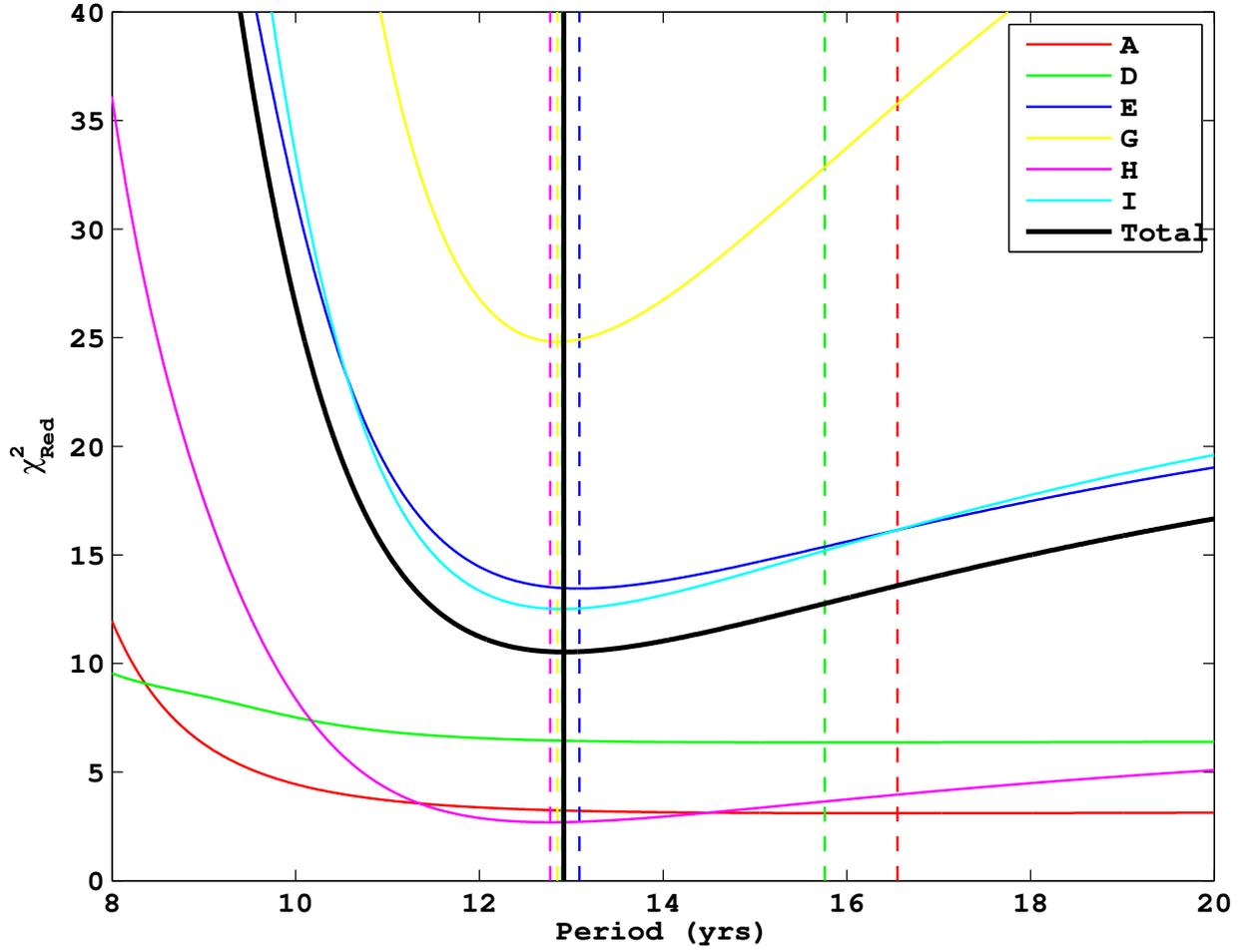}
\caption{Reduced $\chi^2$ as a function of $\Pi$ (see Equation \ref{model}) for each of the coherent $O-C$ diagrams. The average minimum is at $12.9$ years. }\label{pfit}
\end{figure}

\else

\begin{figure} 
\includegraphics[width=1\columnwidth,angle=0]{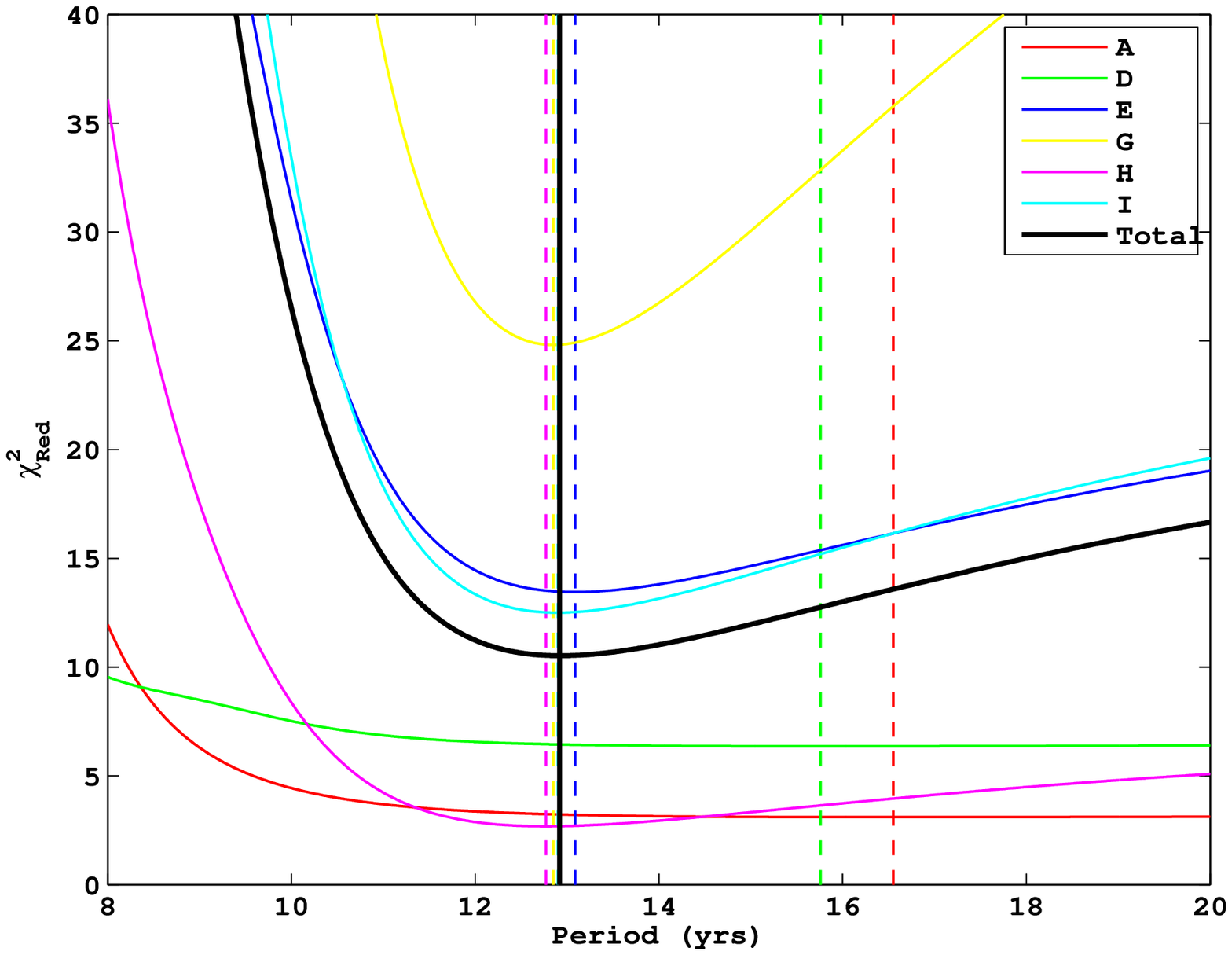}
\caption{Reduced $\chi^2$ as a function of $\Pi$ (see Equation \ref{model}) for each of the coherent $O-C$ diagrams. The average minimum is at $12.9$ years. }\label{pfit}
\end{figure}

\fi

\begin{table}\small
\begin{center}
\caption{Resultant fit parameters from fitting Equation \ref{model} to the six $O-C$ diagrams. Note the high values of reduced $\chi^2$ fit means that some of the formal quoted errors are unreliable, even though the model reproduces the overall trends in the data very well.}\label{fit_table}
\begin{tabular}{|c|c|cccc|}
\tableline
Label & Period [s] & $\dot{P}$ [$10^{-14}$] & $\alpha$ [$10^{-9}\frac{1}{s}$] & $\Phi$ [Deg] & $\chi^2_{Red}$\\
\tableline
A & 134 & -28(1) & 4.0(.1) & 160(2) & 3.3\\
D & 205 & -1(2) & 1.33(.09) & 187(4) & 6.6\\
E & 257 & 13.0(.8) & 2.48(.02) & 0(1)& 14\\
G & 281 & -131(1) & 6.44(.03)& 166.6(.2) & 26\\
H & 287 & 88(6) & 6.8(.1) & 1(1) & 2.8\\
I & 333 & 313(4) & 7.31(.07) & -1(1) & 13\\

\tableline

\end{tabular}
\end{center}
\end{table}

\ifblackandwhite
\begin{figure} 
\includegraphics[width=1\columnwidth,angle=0]{Rbw.eps}
\caption{Residuals of the fit of Equation \ref{model} to the $O-C$ diagrams of D,E,G,H and I. The open circles are the weighted average residuals for each data set while the filled circles are the weighted average of the absolute value of the residuals.}\label{res}
\end{figure}

\else
\begin{figure} 
\includegraphics[width=1\columnwidth,angle=0]{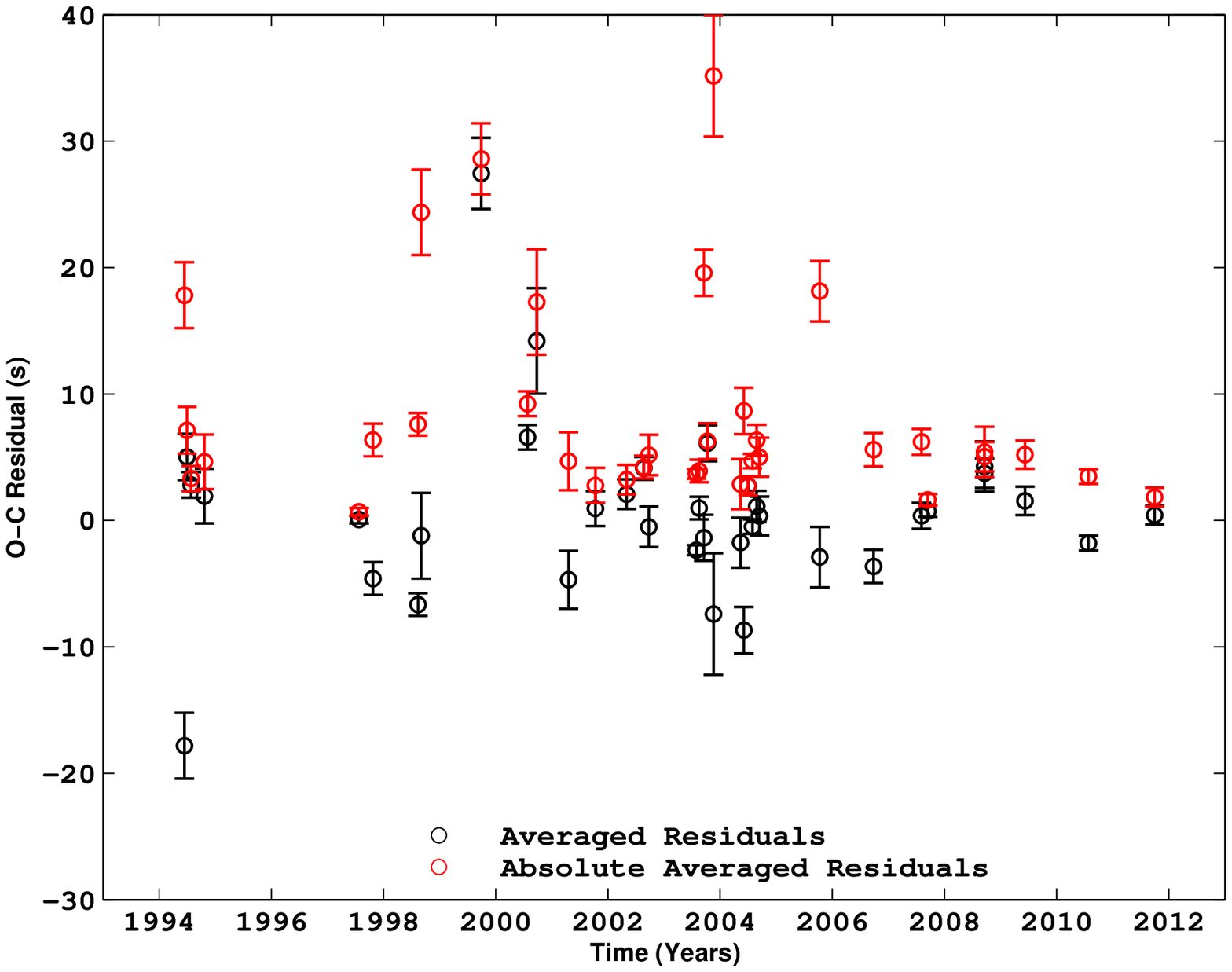}
\caption{Residuals of the fit of Equation \ref{model} to the $O-C$ diagrams of D,E,G,H and I. The black points are the weighted average residuals for each data set while the red points are the weighted average of the absolute value of the residuals.}\label{res}
\end{figure}

\fi

\section{Discussion}
\subsection{Planetary Detection with the Pulsation Timing Method}
The sinusoidal component of the variations shown in Figures \ref{ocA}-\ref{ocI} do not share the same amplitude and phase, so are clearly not due to the effects of a planetary companion. Had only one of the pulsation frequencies in this star been analyzed, this paper could very well be announcing the first detection of a planet around a white dwarf. The discovery that multiple $O-C$ diagrams can show similar periodic behavior when not in the presence of a planetary companion is alarming. Pulsation timing based white dwarf planet detection should now require supplemental confirmation, even when similar pulsation timing variations are observed in multiple pulsation frequencies. Until the physical mechanisms behind these variations are identified as specific to white dwarfs, these observations cast at least some shadow of doubt on the effectiveness of pulsation timing to \textit{reliably} detect planetary companions to subdwarf B pulsators, and raise at least some skepticism as to the existence the subdwarf B planet V391 Pegasi b. We note that this does not affect the ability of the pulsation timing method to \textit{exclude} the existence of planets around white dwarfs and other variables. The lack of known white dwarf planets is gaining statistical significance and further timing observations of pulsating white dwarfs could place strong constraints on post main sequence stellar evolution. 

\subsection{The Combination Frequency A}\label{comb}
\cite{2008MNRAS.387..137S} identified the frequency of A to be an additive combination of the frequencies of E and G, a well known occurrence in white dwarf pulsators (see e.g. \citealt{2009ApJ...693..564P}). Any perturbation to the frequency of E or G should result in an identical perturbation to the frequency of A. The frequency perturbations found by fitting Equation \ref{model} to the $O-C$ diagram of E, G, and A can be compared, see Table \ref{AEG}. The agreement is outstanding. This is strong evidence that Equation \ref{model} is effectively modeling the the frequency perturbations and that the parabolic and sinusoidal components of the perturbation are linearly independent. It is also the most conclusive evidence that combination frequencies are indeed exactly equal to the additive frequency of the parents as predicted by \cite{2001MNRAS.323..248W}.

\begin{table}\small
\begin{center}
\caption{Comparison of the modeled variations of the combination frequency A to its parents E and G.}\label{AEG}
\begin{tabular}{|c|c|c|}
\tableline
Parameter & A & E + G\\
\tableline
$\dot{f} = -\frac{\dot{P}}{P^2}$  [$\frac{10^{-18}}{s^2}$] & 14.6(.2) &  15.3(.5)\\
$\alpha$  [$\frac{10^{-9}}{s}$] & 4.0(.1) & 4.07(.03)\\
$\phi$  [deg]& 160(2) & 158.7(.2) \\

\tableline
\end{tabular}
\end{center}
\end{table}

\subsection{An Asteroseismologic Interpretation of the Results}\label{astero}
The frequency of these pulsations are dependent on many physical quantities, including the star's rotational velocity, magnetic field strength, and magnetic field geometry (for an in depth review, see \citealt{1979nos..book.....U}). Any redistribution of angular momentum or change in magnetic field strength (or geometry) would change the observed frequencies. The sign of the first order frequency perturbation caused by changes in rotation is well known to depend solely on whether the pulsations are moving with or against rotation. In other words, if one pulsation is moving with rotation and another against it, a redistribution of angular momentum would perturb the frequency of the two pulsations in the opposite direction. The variations in pulsation frequencies E, H, and I have positive $\dot{P}$ and $\phi\approx0$ degrees while pulsation frequencies D and G have negative $\dot{P}$ and $\phi$ close to 180 degrees. We speculate that one of these groups of pulsation frequencies could be moving with rotation and the other against rotation and that the observed variations in $O-C$ may be caused by changes in the star's rotation profile. Additionally, the long period g-mode pulsation frequencies in white dwarfs are well known to be more affected by physical properties at the surface of the star than short period g-mode pulsation frequencies. The observed frequency perturbations appear to affect the longer period pulsations more, a hint that the corresponding physical processes responsible for these frequency variations may be near the surface of the star. We again emphasize the speculative nature of these interpretations. The possible effects of magnetic field variations and quantitative modeling of these frequency variations will be investigated in the primary author's doctoral dissertation.

\acknowledgements{JD, JLP, and HLS thank the Crystal Trust for supporting this research. JD thanks the Delaware Space Grant Consortium for financial support. DJS and TS thank the Department of Physics and Astronomy at the University of Canterbury for the generous allocation of telescope time for this project and the VUW Faculty of Science for financial assistance. DK thanks the University of the Western Cape and the South African Foundation for Research Development (FRD) for continuing financial support. This paper uses observations made at the South African Astronomical Observatory (SAAO).}

\bibliography{bibliography}{}

\begin{thebibliography}{28}
\expandafter\ifx\csname natexlab\endcsname\relax\def\natexlab#1{#1}\fi

\bibitem[{{Barlow} {et~al.}(2011){Barlow}, {Dunlap}, {Clemens}, {Reichart},
  {Ivarsen}, {Lacluyze}, {Haislip}, \& {Nysewander}}]{2011MNRAS.414.3434B}
{Barlow}, B.~N., {Dunlap}, B.~H., {Clemens}, J.~C., {Reichart}, D.~E.,
  {Ivarsen}, K.~M., {Lacluyze}, A.~P., {Haislip}, J.~B., \& {Nysewander}, M.~C.
  2011, \mnras, 414, 3434

\bibitem[{{Bischoff-Kim}(2008)}]{2008CoAst.154...16B}
{Bischoff-Kim}, A. 2008, Communications in Asteroseismology, 154, 16

\bibitem[{{Buchler} \& {Koll{\'a}th}(2011)}]{2011ApJ...731...24B}
{Buchler}, J.~R., \& {Koll{\'a}th}, Z. 2011, \apj, 731, 24

\bibitem[{{Dalessio}(2010)}]{2010AAS...21545209D}
{Dalessio}, J. 2010, in Bulletin of the American Astronomical Society, Vol.~42,
  American Astronomical Society Meeting Abstracts 215, 452.09--+

\bibitem[{{Dalessio} {et~al.}(2012){Dalessio}, {Provencal}, {Barlow}, \&
  {Shipman}}]{Dalessio}
{Dalessio}, J., {Provencal}, J.~L., {Barlow}, B.~N., \& {Shipman}, H.~L. 2012,
  submitted, awaiting publication in the proceedings of the \textit{12th
  European Workshop on White Dwarfs.}

\bibitem[{{Eastman} {et~al.}(2010){Eastman}, {Siverd}, \&
  {Gaudi}}]{2010PASP..122..935E}
{Eastman}, J., {Siverd}, R., \& {Gaudi}, B.~S. 2010, \pasp, 122, 935

\bibitem[{{Hermes} {et~al.}(2010){Hermes}, {Mullally}, {Winget}, {Montgomery},
  {Miller}, \& {Ellis}}]{2010AIPC.1273..446H}
{Hermes}, J.~J., {Mullally}, F., {Winget}, D.~E., {Montgomery}, M.~H.,
  {Miller}, G.~F., \& {Ellis}, J.~L. 2010, in American Institute of Physics
  Conference Series, Vol. 1273, American Institute of Physics Conference
  Series, ed. {K.~Werner \& T.~Rauch}, 446--449

\bibitem[{{Hermes} {et~al.}(2012){Hermes}, {Kilic}, {Brown}, {Winget}, {Allende
  Prieto}, {Gianninas}, {Mukadam}, {Cabrera-Lavers}, \&
  {Kenyon}}]{2012ApJ...757L..21H}
{Hermes}, J.~J., {et~al.} 2012, \apjl, 757, L21

\bibitem[{{Holman} {et~al.}(2010){Holman}, {Fabrycky}, {Ragozzine}, {Ford},
  {Steffen}, {Welsh}, {Lissauer}, {Latham}, {Marcy}, {Walkowicz}, {Batalha},
  {Jenkins}, {Rowe}, {Cochran}, {Fressin}, {Torres}, {Buchhave}, {Sasselov},
  {Borucki}, {Koch}, {Basri}, {Brown}, {Caldwell}, {Charbonneau}, {Dunham},
  {Gautier}, {Geary}, {Gilliland}, {Haas}, {Howell}, {Ciardi}, {Endl},
  {Fischer}, {F{\"u}r{\'e}sz}, {Hartman}, {Isaacson}, {Johnson}, {MacQueen},
  {Moorhead}, {Morehead}, \& {Orosz}}]{2010Sci...330...51H}
{Holman}, M.~J., {et~al.} 2010, Science, 330, 51

\bibitem[{{Kepler} {et~al.}(2005){Kepler}, {Costa}, {Castanheira}, {Winget},
  {Mullally}, {Nather}, {Kilic}, {von Hippel}, {Mukadam}, \&
  {Sullivan}}]{2005ApJ...634.1311K}
{Kepler}, S.~O., {et~al.} 2005, \apj, 634, 1311

\bibitem[{{Koen} {et~al.}(1995){Koen}, {O'Donoghue}, {Stobie}, {Kilkenny}, \&
  {Ashley}}]{1995MNRAS.277..913K}
{Koen}, C., {O'Donoghue}, D., {Stobie}, R.~S., {Kilkenny}, D., \& {Ashley}, R.
  1995, \mnras, 277, 913

\bibitem[{{Mullally} {et~al.}(2009){Mullally}, {Reach}, {De Gennaro}, \&
  {Burrows}}]{2009ApJ...694..327M}
{Mullally}, F., {Reach}, W.~T., {De Gennaro}, S., \& {Burrows}, A. 2009, \apj,
  694, 327

\bibitem[{{Mullally} {et~al.}(2008){Mullally}, {Winget}, {De Gennaro},
  {Jeffery}, {Thompson}, {Chandler}, \& {Kepler}}]{2008ApJ...676..573M}
{Mullally}, F., {Winget}, D.~E., {De Gennaro}, S., {Jeffery}, E., {Thompson},
  S.~E., {Chandler}, D., \& {Kepler}, S.~O. 2008, \apj, 676, 573

\bibitem[{{Oksala} {et~al.}(2012){Oksala}, {Wade}, {Townsend}, {Owocki},
  {Kochukhov}, {Neiner}, {Alecian}, \& {Grunhut}}]{2012MNRAS.419..959O}
{Oksala}, M.~E., {Wade}, G.~A., {Townsend}, R.~H.~D., {Owocki}, S.~P.,
  {Kochukhov}, O., {Neiner}, C., {Alecian}, E., \& {Grunhut}, J. 2012, \mnras,
  419, 959

\bibitem[{{Provencal} {et~al.}(2009){Provencal}, {Montgomery}, {Kanaan},
  {Shipman}, {Childers}, {Baran}, {Kepler}, {Reed}, {Zhou}, {Eggen}, {Watson},
  {Winget}, {Thompson}, {Riaz}, {Nitta}, {Kleinman}, {Crowe}, {Slivkoff},
  {Sherard}, {Purves}, {Binder}, {Knight}, {Kim}, {Chen}, {Yang}, {Lin}, {Lin},
  {Chen}, {Jiang}, {Sergeev}, {Mkrtichian}, {Andreev}, {Janulis}, {Siwak},
  {Zola}, {Koziel}, {Stachowski}, {Paparo}, {Bognar}, {Handler}, {Lorenz},
  {Steininger}, {Beck}, {Nagel}, {Kusterer}, {Hoffman}, {Reiff}, {Kowalski},
  {Vauclair}, {Charpinet}, {Chevreton}, {Solheim}, {Pakstiene}, {Fraga}, \&
  {Dalessio}}]{2009ApJ...693..564P}
{Provencal}, J.~L., {et~al.} 2009, \apj, 693, 564

\bibitem[{{Silvotti} {et~al.}(2007){Silvotti}, {Schuh}, {Janulis}, {Solheim},
  {Bernabei}, {{\O}stensen}, {Oswalt}, {Bruni}, {Gualandi}, {Bonanno},
  {Vauclair}, {Reed}, {Chen}, {Leibowitz}, {Paparo}, {Baran}, {Charpinet},
  {Dolez}, {Kawaler}, {Kurtz}, {Moskalik}, {Riddle}, \&
  {Zola}}]{2007Natur.449..189S}
{Silvotti}, R., {et~al.} 2007, \nat, 449, 189

\bibitem[{{Stumpff}(1980)}]{1980A&AS...41....1S}
{Stumpff}, P. 1980, \aaps, 41, 1

\bibitem[{{Sullivan}(2000)}]{2000BaltA...9..425S}
{Sullivan}, D.~J. 2000, Baltic Astronomy, 9, 425

\bibitem[{{Sullivan}(2005)}]{2005ASPC..334..495S}
{Sullivan}, D.~J. 2005, in Astronomical Society of the Pacific Conference
  Series, Vol. 334, 14th European Workshop on White Dwarfs, ed. {D.~Koester \&
  S.~Moehler}, 495--+

\bibitem[{{Sullivan} {et~al.}(2007){Sullivan}, {Metcalfe}, {O'Donoghue},
  {Winget}, {Kilkenny}, {van Wyk}, {Kanaan}, {Kepler}, {Nitta}, {Kawaler},
  {Montgomery}, {Nather}, {Steeghs}, {Koester}, {Bergeron}, {O'Brien}, {Wood},
  {Jiang}, {Leibowitz}, {Ibbetson}, {Zola}, {Krzesinski}, {Pajdosz},
  {Vauclair}, {Dolez}, \& {Chevreton}}]{2007ASPC..372..629S}
{Sullivan}, D.~J., {et~al.} 2007, in Astronomical Society of the Pacific
  Conference Series, Vol. 372, 15th European Workshop on White Dwarfs, ed.
  R.~{Napiwotzki} \& M.~R. {Burleigh}, 629

\bibitem[{{Sullivan} {et~al.}(2008){Sullivan}, {Metcalfe}, {O'Donoghue},
  {Winget}, {Kilkenny}, {van Wyk}, {Kanaan}, {Kepler}, {Nitta}, {Kawaler},
  {Montgomery}, {Nather}, {O'Brien}, {Bischoff-Kim}, {Wood}, {Jiang},
  {Leibowitz}, {Ibbetson}, {Zola}, {Krzesinski}, {Pajdosz}, {Vauclair},
  {Dolez}, \& {Chevreton}}]{2008MNRAS.387..137S}
{Sullivan}, D.~J., {et~al.} 2008, \mnras, 387, 137

\bibitem[{{Taylor} {et~al.}(1979){Taylor}, {Fowler}, \&
  {McCulloch}}]{1979Natur.277..437T}
{Taylor}, J.~H., {Fowler}, L.~A., \& {McCulloch}, P.~M. 1979, \nat, 277, 437

\bibitem[{{Thompson} \& {Mullally}(2009)}]{2009JPhCS.172a2081T}
{Thompson}, S.~E., \& {Mullally}, F. 2009, Journal of Physics Conference
  Series, 172, 012081

\bibitem[{{Thorsett} {et~al.}(1993){Thorsett}, {Arzoumanian}, \&
  {Taylor}}]{1993ApJ...412L..33T}
{Thorsett}, S.~E., {Arzoumanian}, Z., \& {Taylor}, J.~H. 1993, \apjl, 412, L33

\bibitem[{{Unno} {et~al.}(1979){Unno}, {Osaki}, {Ando}, \&
  {Shibahashi}}]{1979nos..book.....U}
{Unno}, W., {Osaki}, Y., {Ando}, H., \& {Shibahashi}, H. 1979, {Nonradial
  oscillations of stars}

\bibitem[{{Winget} {et~al.}(2004){Winget}, {Sullivan}, {Metcalfe}, {Kawaler},
  \& {Montgomery}}]{2004ApJ...602L.109W}
{Winget}, D.~E., {Sullivan}, D.~J., {Metcalfe}, T.~S., {Kawaler}, S.~D., \&
  {Montgomery}, M.~H. 2004, \apjl, 602, L109

\bibitem[{{Wolszczan} \& {Frail}(1992)}]{1992Natur.355..145W}
{Wolszczan}, A., \& {Frail}, D.~A. 1992, \nat, 355, 145

\bibitem[{{Wu}(2001)}]{2001MNRAS.323..248W}
{Wu}, Y. 2001, \mnras, 323, 248

\end{thebibliography}
\bibliographystyle{apj}

\end{document}